\documentclass[12pt]{article}
\catcode`\@=11
\@addtoreset{equation}{section}

\global\arraycolsep=2pt 
\newcommand{\ga}{\gamma}

\newcommand{\Tr}{{\rm Tr}}
\newcommand{\str}{{\rm STr}}

\newcommand*{\AdS}[1]{\ensuremath{\text{AdS}_{#1}}}

\newcommand{\no}{\nonumber}

\newcommand{\bd}{{\rm d}}

\usepackage{mathrsfs,amsbsy,amssymb,latexsym,amsfonts,amsmath,cite}
\usepackage{graphicx,color}
\usepackage{mathtools}

\allowdisplaybreaks

\begin{document}

\begin{flushright}
\parbox{4cm}
{KUNS-2657}
\end{flushright}

\vspace*{1.5cm}

\begin{center}
{\Large \bf
Yang-Baxter deformations of $W_{2,4}\times T^{1,1}$ \\ 
and the associated T-dual models }
\vspace*{1.5cm}\\
{\large Jun-ichi Sakamoto\footnote{E-mail:~sakajun@gauge.scphys.kyoto-u.ac.jp}
and Kentaroh Yoshida\footnote{E-mail:~kyoshida@gauge.scphys.kyoto-u.ac.jp}}
\end{center}

\vspace*{0.25cm}

\begin{center}
{\it Department of Physics, Kyoto University, \\
Kitashirakawa, Kyoto 606-8502, Japan}
\end{center}

\vspace{2cm}

\begin{abstract}
Recently, for principal chiral models and symmetric coset sigma models, Hoare and Tseytlin 
proposed an interesting conjecture that the Yang-Baxter deformations with the homogeneous 
classical Yang-Baxter equation are equivalent to non-abelian T-dualities with topological terms. 
It is significant to examine this conjecture for non-symmetric (i.e., non-integrable) cases. 
Such an example is the $W_{2,4}\times T^{1,1}$ background.  
In this note, we study Yang-Baxter deformations of type IIB string theory defined 
on $W_{2,4}\times T^{1,1}$ and the associated T-dual models, and show that 
this conjecture is valid even for this case. Our result indicates that the conjecture 
would be valid beyond integrability. 
\end{abstract}

\setcounter{footnote}{0}
\setcounter{page}{0}
\thispagestyle{empty}

\newpage

\tableofcontents

\section{Introduction}

A prototypical example of the AdS/CFT correspondence \cite{M} is a duality 
between type IIB string theory on the AdS$_5\times$S$^5$ background\footnote{This theory 
is often abbreviated as the AdS$_5\times$S$^5$ superstring.} 
and the four-dimensional $\mathcal{N}=4$ $SU(N)$ super Yang-Mills (SYM) theory 
in the large $N$ limit. As a remarkable feature, an integrable structure exists 
behind this correspondence \cite{review}. On the string-theory side, 
it is well known that the classical action of the AdS$_5\times$S$^5$ superstring \cite{MT} 
enjoys the $\mathbb{Z}_4$-grading and it ensures the classical integrability in the sense of 
kinematical integrability \cite{BPR} (For nice reviews on this issue, see 
\cite{AF-review,review1,Borsato}). 

\medskip 

One of the fascinating subjects on this integrability is to study Yang-Baxter (YB) deformations 
\cite{Klimcik,DMV1,MY} of the AdS$_5\times$S$^5$ superstring \cite{DMV2,KMY}. 
YB deformations were originally proposed by Klimcik \cite{Klimcik} 
for principal chiral models with the modified classical Yang-Baxter equation (mCYBE). 
Those were then generalized to symmetric coset sigma models \cite{DMV1} 
and the homogeneous CYBE. For affine symmetries related to the deformed models, 
see \cite{KMY-alg,KOY,KY-Sch}. 

\medskip 

We are concerned here with the YB deformations with the homogeneous CYBE \cite{KMY,MY}. 
The YB deformed AdS$_5\times$S$^5$ backgrounds have been intensively 
studied in the recent progress \cite{LM-MY,MR-MY,Sch-MY,KMY-SUGRA,MY-duality,MY-summary,
Stijn1,Stijn2,KKSY,KY,HvT,ORSY,BW,OvT,Stijn3}. 
A remarkable progress is the discovery of the unimodularity condition \cite{BW}, 
under which the deformed spacetime satisfies the on-shell condition of type IIB supergravity. 
This unimodular class includes all of the abelian classical $r$-matrices. 
A series of works \cite{LM-MY,MR-MY,Sch-MY,KY} have identified the abelian classical $r$-matrices 
associated with $\gamma$-deformations of S$^5$ \cite{LM,Frolov}, 
gravity duals of non-commutative gauge theories \cite{HI,MR}
and Schr$\ddot{\rm o}$dinger spacetimes \cite{MMT}. 

\medskip 

On the other hand, if a classical $r$-matrix does not satisfy the unimodularity condition, then 
the resulting background is not a solution of type IIB supergravity, 
but satisfies the generalized equations 
of motion \cite{AFHRT} (as supported by a series of works \cite{KY,HvT,ORSY,BW}). 
The appearance of the generalized type IIB supergravity is rather inevitable 
because the generalized equations are reproduced from the kappa-symmetry constraints 
of the Green-Schwarz string theories on arbitrary backgrounds \cite{WT}\footnote{Note here that 
this is a new result obtained recently, while it has been well known that 
the on-shell condition of type IIB supergravity leads to the kappa-invariant Green-Schwarz 
string theories \cite{Townsend}.} (though those were discovered so as to support 
the $\eta$-deformed background \cite{ABF,ABF2} as a solution). 
Solutions of the generalized supergravity can be mapped to solutions of the usual supergravities 
via ``T-dualities'' \cite{AFHRT,HT-sol} along non-isometric directions. 
Recently, the modified double field theory description 
has been constructed in \cite{SUY} as the underlying structure behind the generalized gravities. 
By following it, the ``T-dualities'' can be naturally understood as $O(D,D)$ transformations. 
As yet another approach is a direct derivation 
from the (non-modified) exceptional field theory \cite{Magro}. 

\medskip 

Recently, for the homogeneous CYBE case, Hoare and Tseytlin proposed an interesting 
conjecture that the YB deformations are equivalent to non-abelian T-dualities 
for principal chiral models and coset sigma models \cite{HT-conjecture}. 
Then a proof of this conjecture was provided in \cite{BW-proof}. 
This equivalence would be very important because it is helpful in studying what happens 
to the string target spacetime, or what happens to the gauge-theory side after performing 
YB deformations. For the recent progress along this line, see \cite{Thompson}. 

\medskip 

As a possible generalization, it is also interesting to examine YB deformations of 
non-integrable homogeneous backgrounds\footnote{In order to perform YB deformations, 
a coset representation of the target space is necessary. Hence the homogeneity is supposed here.}. 
Such an example is the AdS$_5\times T^{1,1}$ background, where 
$T^{1,1}$ is a five-dimensional Sasaki-Einstein manifold \cite{CD}. 
This background was originally introduced by Klebanov and Witten \cite{KW} as a gravity 
dual of a superconformal field theory in four dimensions. This $T^{1,1}$ is known 
as a non-integrable background because classical string solutions on $T^{1,1}$ 
exhibit the chaotic behavior \cite{BZ}. 
On the other hand, YB deformations of $T^{1,1}$ are studied in \cite{CMY} 
and TsT transformations of $T^{1,1}$ \cite{LM,CO} can be reproduced as YB deformations. 
This result indicates that YB deformations would work well beyond integrability, 
although those were originally proposed as integrable deformations. 

\medskip 

Along the above line, it would be nice to study the Hoare-Tseytlin conjecture for non-integrable 
cases. However, the $T^{1,1}$ background is compact and hence the conjecture would not be 
so non-trivial because the YB deformations with the homogeneous CYBE become abelian 
and always satisfy the unimodularity condition. To expand our argument so as to include 
non-unimodular cases, it is better to study a non-integrable, non-compact 
and homogeneous space. Such an example is an Einstein manifold $W_{2,4}$ 
(which is a non-symmetric coset space).    
The $W_{2,4} \times$S$^5$ background is introduced in \cite{BPSV} 
to study a holographic principle. 
Classical chaotic string solutions have not been constructed explicitly on the $W_{2,4}$ space. 
However, the $W_{2,4}$ geometry should be non-integrable 
because it can be realized as a double Wick rotation of $T^{1,1}$\,. 
Thus $W_{2,4}$ is suitable for our purpose. 

\medskip 

In this note, we will argue the Hoare-Tseytlin conjecture for the $W_{2,4} \times T^{1,1}$
background\footnote{A similar background is studied in \cite{PZT}, 
but different from the one we are concerned with here. 
It contains an NS-NS two-form but no R-R flux, and it is a consistent NS-NS string 
background. The model of \cite{PZT} is conformal and one may construct integrable 
deformations of it. The resulting deformed theory is a special case of the theories 
presented in \cite{Sfetsos-new}. We are grateful to A.~Tseytlin, G.~Georgiou and K.~Sfetsos 
for this point.}\,.
We study YB deformations of type IIB string theory defined 
on $W_{2,4}\times T^{1,1}$ and the associated T-dual models, and show that 
this conjecture is valid for this case as well. Our result indicates that the conjecture 
would be valid beyond integrability. 

\medskip 

This note is organized as follows. Section 2 introduces a coset construction of 
the $W_{2,4} \times T^{1,1}$ spacetime. Section 3 gives a short review of the Hoare-Tseytlin 
conjecture for principal chiral models by following the work \cite{BW-proof}.  
In section 4, we consider non-abelian T-dualities and YB deformations of $W_{2,4} \times T^{1,1}$\,. 
Section 5 provides examples of classical $r$-matrices and the associated T-dual models. 
Section 6 is devoted to conclusion and discussion.

\section{Coset construction of $W_{2,4}\times T^{1,1}$}

In this section, we shall introduce the $W_{2,4}\times T^{1,1}$ spacetime. 
This geometry is homogeneous and its metric can be derived by performing a coset construction. 
We describe the procedure of the coset construction in detail. 
This section basically follows the preceding work on $T^{1,1}$ \cite{CMY}, 
but includes a generalization to $W_{2,4}$\,. 

\subsection{The geometries of $W_{2,4}\times T^{1,1}$}

To begin with, we briefly describe the geometry of $W_{2,4}\times T^{1,1}$\,. 

\medskip

Let us first see the $T^{1,1}$ part. The internal space $T^{1,1}$ is a five-dimensional 
Sasaki-Einstein manifold and can be viewed as a 
$U(1)_{\rm R}$-fibration over $SU(2)_{\rm A}\times SU(2)_{\rm B}$ \cite{CO}.
The geometry is equipped with the metric  
\begin{eqnarray}
\bd s^2_{T^{1,1}}&=&\frac{1}{9}\left(\bd\psi+\cos \theta_1\,\bd\phi_1
+\cos \theta_2\,\bd\phi_2\right)^2\no \\
&&+\frac{1}{6}\left[(\bd \theta_1)^2+\sin^2 \theta_1 (\bd\phi_1)^2
+(\bd \theta_2)^2+\sin^2 \theta_2 (\bd\phi_2)^2\right]\,,
\label{Tmetric}
\end{eqnarray}
where $0\leq\theta_1\,, \theta_2<\pi$\,, $0\leq\phi_1\,, \phi_2<2\pi$ and $0\leq\psi<4\pi$\,.
The coordinate $\psi$ parametrizes the $U(1)_{\rm R}$ fiber.
The isometry group is $SU(2)_{\rm A} \times SU(2)_{\rm B} \times U(1)_{\rm R}$\,.
It has been revealed that the $T^{1,1}$ manifold is represented by a coset \cite{CMY}
\begin{eqnarray}
T^{1,1}=\frac{SU(2)_{\rm A}\times SU(2)_{\rm B} \times U(1)_{\rm R}}{U(1)_{\rm A}
\times U(1)_{\rm B}}\,.
\label{Tcoset}
\end{eqnarray}
Here $SU(2)_{\rm A} \times SU(2)_{\rm B}$ and $U(1)_{\rm R}$ correspond to  
a flavor symmetry and an $R$-symmetry in the dual superconformal field theory \cite{KW}, 
respectively. Because the coset (\ref{Tcoset}) is not symmetric, the classical integrability of 
a non-linear sigma model in two dimensions with target space $T^{1,1}$ is not ensured automatically. 
Indeed, chaotic string solutions are presented in \cite{BZ} and hence 
the sigma model is shown to be non-integrable. 

\medskip

The next is to see the Lorentzian manifold $W_{2,4}$\,.
This is also a five-dimensional Einstein space with the metric 
\begin{eqnarray}
\bd s^2_{W_{2,4}} &=& 
-\frac{1}{9}\left(\bd\chi+\cosh y_1\,\bd\psi_1+\cosh y_2\,\bd\psi_2\right)^2\no \\
&&
+\frac{1}{6}\left[(\bd y_1)^2+\sinh^2 y_1 (\bd\psi_1)^2
+(\bd y_2)^2+\sinh^2 y_2 (\bd\psi_2)^2\right]\,.
\label{Wmetric}
\end{eqnarray}
Here $0\leq y_1\,, y_2<\infty$\,, $0\leq\psi_1\,, \psi_2<2\pi$ and $0\leq\chi<4\pi$\,. 
The $W_{2,4}$ geometry can also be regarded as a $U(1)$-fibration over 
${\rm E}\AdS{2}\times {\rm E}\AdS{2}$ and then the coordinate $\chi$ parametrizes 
the $U(1)$ fiber. 

\medskip 

Note here that the $W_{2,4}$ metric can be derived formally by performing a double Wick rotation
\begin{eqnarray}
\theta_{1\,,2}\to i y_{1\,,2}\,,\qquad \psi\to \chi\,,\qquad \phi_{1\,,2}\to \psi_{1\,,2}\,,
\label{dW}
\end{eqnarray}
for the $T^{1,1}$ metric (up to the overall sign). Hence   
the geometry of $W_{2,4}$ is represented by the following coset:\footnote{In \cite{BPSV}, 
a coset $SO(2,2)/SO(2)$ is argued to describe $W_{2,4}$\,. However, this coset does not work 
in performing the coset construction. This is the case for $T^{1,1}$ as well. The popular coset 
$(SU(2) \times SU(2))/U(1)$ does not work for the coset construction. This point is explicitly 
denoted in the seminal paper \cite{CD}. The proper coset (\ref{Tcoset}) and the supertrace operation 
have been clarified in \cite{CMY}.}
\begin{eqnarray}
W_{2,4}=\frac{SL(2)_{\rm a}\times SL(2)_{\rm b} \times SO(2)_{\rm r}}{
U(1)_{\rm a}\times U(1)_{\rm b}}\,.
\label{Wcoset}
\end{eqnarray}
The coset (\ref{Wcoset}) is not symmetric as well. 
Chaotic string solutions have not been constructed explicitly on the $W_{2,4}$ background.  
However, $W_{2,4}$ should also be non-integrable like $T^{1,1}$ 
because $W_{2,4}$ can be realized as a double Wick rotation of $T^{1,1}$\,, as denoted above.

\subsection{Coset construction of $W_{2,4}\times T^{1,1}$}

Let us next derive the metric of $W_{2,4}\times T^{1,1}$ by performing a coset construction 
explicitly with the following coset: 
\begin{eqnarray}
W_{2,4}\times T^{1,1}=\frac{SL(2)_{\rm a}\times SL(2)_{\rm b} \times SO(2)_{\rm r}}{
U(1)_{\rm a}\times U(1)_{\rm b}}\times \frac{SU(2)_{\rm A}\times SU(2)_{\rm B} 
\times U(1)_{\rm R}}{U(1)_{\rm A}\times U(1)_{\rm B}}\,.
\label{WTcoset}
\end{eqnarray}
The derivation of the $T^{1,1}$ metric is just a review of \cite{CMY}, 
but that of the $W_{2,4}$ one has not been presented yet. 

\medskip

It is convenient to introduce a matrix representation of the generators. 
We will take the $SU(2)$ generators $F_i~(i=1,2,3)$  
and the $SL(2)$ generators $L_{\mu}~(\mu=0,1,2)$ as follows:  
\begin{eqnarray}
F_1 &=& -\frac{i}{2}\sigma_1\,, \qquad F_2 = -\frac{i}{2}\sigma_2\,, \qquad 
F_3 = -\frac{i}{2}\sigma_3\,,  
\no \\
L_0&=& \frac{i}{2}\sigma_3\,, 
%-F_3\,, 
\qquad~~ L_1= -\frac{1}{2}\sigma_2\,, 
%-i F_2 \,,
\qquad 
L_2= \frac{1}{2}\sigma_1\,.
%i F_1\,. 
\end{eqnarray}
Here $\sigma_a~(a=1,2,3)$ are the standard Pauli matrices. 

\medskip 

By following the procedure presented in \cite{CMY},
we choose the fundamental representations of $(5|5)\times (5|5)$ supermatrix 
rather than the bosonic $10\times 10$ matrices.
We take $L_{\mu}^1\,,L_{\mu}^2$ and $K$ as the generators of the Lie algebras 
$\mathfrak{sl}(2)_{\rm a}$\,, $\mathfrak{sl}(2)_{\rm b}$ and $\mathfrak{so}(2)$\,, respectively.
Then a matrix realization of the generators is given by
\begin{align}
L^1_\mu=
\left( \hspace{-\arraycolsep}
\begin{array}{c|@{}c}
\hspace{-1pt} $\smash{\lower 2ex\hbox{$0_{5\times5}$}}$\hspace{2pt} 
&\hspace{5pt}$\smash{\lower 2ex\hbox{$0_{5\times5}$}}$\\[15pt] \hline
\hspace{-1pt} $\smash{\lower 0ex\hbox{$0_{5\times5}$}}$\hspace{2pt} &
\begin{array}{cc}
\,L_{\mu} &
\begin{array}{ccc}
0 & 0 & 0 \\
0 & 0 & 0 
\end{array}
\\
\begin{array}{cc}
0 & 0 \\ 
0 & 0 \\ 
0 & 0
\end{array} 
& \begin{array}{ccc}
0 & 0 & 0 \\ 
0 & 0 & 0\\ 
0 & 0 & 0
\end{array}
\end{array}
\end{array}
\right)\,, ~~
L^2_\mu=
\left( \hspace{-\arraycolsep}
\begin{array}{c|c}
\hspace{-1pt}$\smash{\lower 2ex\hbox{$0_{5\times5}$}}$\hspace{2pt}
&$\smash{\lower 2ex\hbox{$0_{5\times5}$}}$\\[15pt] 
\hline
\hspace{-1pt} $\smash{\lower 0ex\hbox{$0_{5\times5}$}}$\hspace{2pt}  &
\begin{array}{ccc}
\begin{array}{cc}
0 & 0 \\ 
0 & 0 
\end{array}
&
\begin{array}{cc}
0 & 0 \\ 
0 & 0 
\end{array}
& 
\begin{array}{c}
0 \\ 
0 
\end{array}
\\
\begin{array}{cc}
0 & 0 \\ 
0 & 0 
\end{array}
&L_{\mu}&
\begin{array}{c}
0 \\ 
0 
\end{array}
\\
\begin{array}{cc}
0 & 0 
\end{array}
& \begin{array}{cc}
0 & 0  
\end{array}
&0\\
\end{array}
\end{array}
\right)\,, ~~
K=-\frac{i}{2}
\left( \hspace{-\arraycolsep}
\begin{array}{c|c}
\begin{array}{ccccc}
1 & 0 & 0 & 0 & 0 \\
0 & 0 & 0 & 0 & 0 \\ 
0 & 0 & 0 & 0 & 0 \\ 
0 & 0 & 0 & 0 & 0 \\
0 & 0 & 0 & 0 & 0 
\end{array}
&\hspace{5pt}$\smash{\lower 0ex\hbox{$0_{5\times5}$}}$ \\
\hline
\hspace{5pt}$\smash{\lower 2ex\hbox{$0_{5\times5}$}}$\hspace{5pt}   &
\hspace{5pt}$\smash{\lower 2ex\hbox{$0_{5\times5}$}}$    \\
&
\end{array}
\right)\,.
\label{sup-matW}
\end{align}
Similarly, $F_i^1\,,F_i^2$ and $M$ are the generators of  the Lie algebras 
$\mathfrak{su}(2)_{\rm A}$\,,  $\mathfrak{su}(2)_{\rm B}$ and $\mathfrak{u}(1)_{\rm R}$\,, 
respectively. These are represented by the following matrices: 
\begin{align}
F^1_i=
\left( \hspace*{-5pt}
%\hspace{-\arraycolsep}
\begin{array}{c|@{}c}
\begin{array}{ccc}
\begin{array}{cc}
0 & 0 \\ 
0 & 0
\end{array}
&
\begin{array}{cc}
0 & 0 \\ 
0 & 0
\end{array}
&
\begin{array}{c}
0 \\ 
0 
\end{array} 
\\
\begin{array}{cc}
0 & 0 \\ 
0 & 0
\end{array}
&F_i&
\begin{array}{c}
0 \\ 
0 
\end{array}
\\
\begin{array}{cc}
0 & 0 
\end{array}
&
\begin{array}{cc}
0 & 0 
\end{array}
&0
\end{array}
&\hspace{5pt}$\smash{\lower 0ex\hbox{$0_{5\times5}$}}$\\
%[15pt]
\hline
\hspace{5pt}$\smash{\lower 2ex\hbox{$0_{5\times5}$}}$\hspace{2pt} &
\hspace{5pt}$\smash{\lower 2ex\hbox{$0_{5\times5}$}}$
%\hspace{2pt}
\\ 
&
\end{array}
\right)\,,~~
F^2_i=
\left( \hspace*{-5pt}
%\hspace{-\arraycolsep}
\begin{array}{c|@{}c}
\begin{array}{cc}
\begin{array}{ccc}
0 & 0 & 0 \\
0 & 0 & 0 \\
0 & 0 & 0 
\end{array}
&  
\begin{array}{cc}
0 & 0  \\
0 & 0 \\
0 & 0  
\end{array} 
\\
\begin{array}{ccc}
0 & 0 & 0 \\
0 & 0 & 0 
\end{array} 
&F_i 
\end{array} 
&
\hspace{5pt}
$\smash{\lower 1ex\hbox{$0_{5\times5}$}}$\\
\hline
\hspace{5pt}
$\smash{\lower 2ex\hbox{$0_{5\times5}$}}$&
\hspace{5pt}
$\smash{\lower 2ex\hbox{$0_{5\times5}$}}$\\
&
\end{array}
\right)\,, 
~~M=-\frac{i}{2}
\left( \hspace*{-5pt}
%\hspace{-\arraycolsep}
\begin{array}{c|c}
&\\
%\hspace{5pt}
$
\smash{\lower -2ex\hbox{$0_{5\times5}$}}$
%\hspace{5pt} 
&
%\hspace{5pt}
$
\smash{\lower -2ex\hbox{$0_{5\times5}$}}$ \\
\hline
%\hspace{5pt}
$
\smash{\lower 0ex\hbox{$0_{5\times5}$}}$
%\hspace{5pt}   
& 
\begin{array}{ccccc}
0 & 0 & 0 & 0 & 0 \\
0 & 0 & 0 & 0 & 0 \\
0 & 0 & 0 & 0 & 0 \\
0 & 0 & 0 & 0 & 0 \\
0 & 0 & 0 & 0 & 1 
\end{array}
\end{array}
\right)\,.
\label{sup-matT}
\end{align}
The non-vanishing commutation relations are given by 
\begin{eqnarray}
[L^m_1,L^n_\pm]=\pm\delta^{mn} L^m_\pm\,,\quad~~
[L^m_+,L^n_-]=2\delta^{mn}L^m_1\,,\quad~~
[F^m_i,F^n_j]=\delta^{mn}\epsilon_{ijk}F^m_k\,,
\end{eqnarray}
where $L^m_\pm=L_2^m\pm L^m_0$\,. 
It is helpful to use the following supertrace formulae: 
\begin{eqnarray}
\str(L^m_{\mu} L^n_{\nu}) &=& -\frac{1}{2}\delta^{mn}\eta_{\mu\nu}\,,\quad
\str(F^m_iF^n_j)=-\frac{1}{2}\delta^{mn}\delta_{ij}\,,\no\\
\str(KK)&=&-\str(MM)=-\frac{1}{4}\,.
\end{eqnarray}
Here $\eta_{\mu\nu} \equiv {\rm diag}(-++)$\,. 

\medskip 

To parametrize a representative of the coset (\ref{WTcoset})\,, 
let us introduce here the following orthogonal basis :
\begin{eqnarray}
{\rm Span}_{\mathbb{R}}\{L_0^m\,,L_1^m\,,W\,,F_1^m\,,F_2^m\,,H\}\,,\qquad (m=1,2)\,. 
\label{cosetsp}
\end{eqnarray}
Here $W$ and $H$ are defined as
\begin{eqnarray}
W \equiv L_0^1-L_0^2+K\,,\qquad
H \equiv F_3^1-F_3^2+M\,.
\end{eqnarray}
The denominator of the coset (\ref{WTcoset}) is then spanned 
by the following abelian generators
\begin{eqnarray}
T_1&=&L_0^1+L_0^2\,,\qquad T_2=L_0^1-L_0^2+4 K\,, \no \\
T_3&=&F_3^1+F_3^2\,,\qquad T_4=F_3^1-F_3^2+4 M\,. \label{denominator}
\end{eqnarray}

\medskip

The metric of $W_{2,4}\times T^{1,1}$ can be reproduced by using a representative $g$
of the coset (\ref{WTcoset})\,. 
Then $g$ is decomposed into the $W_{2,4}$ and $T^{1,1}$ parts like
\begin{eqnarray}
g=g_{W_{2,4}}\cdot g_{T^{1,1}}\,,
\end{eqnarray}
where $g_{W_{2,4}}$ and $g_{T^{1,1}}$ are parametrized as 
\begin{eqnarray}
g_{W_{2,4}}&=&\exp\left[\psi_1 L_0^1+\psi_2 L_0^2+2 \chi K\right]
\exp\left[(y_1-i \pi)L_1^1+y_2 L_1^2\right]\,,\label{globalW}\\
g_{T^{1,1}}&=&\exp\left[\phi_1 F_3^1+\phi_2 F_3^2+2 \psi M\right]
\exp\left[(\theta_1+\pi)F_2^1+\theta_2 F_2^2\right]\,.
\label{globalT}
\end{eqnarray}
By performing a coset construction with a left-invariant one-form 
\[
A \equiv -g^{-1}{\rm d}g\,, 
\] 
the metric of $W_{2,4}\times T^{1,1}$\,, which is a sum of  (\ref{Wmetric}) and (\ref{Tmetric})\,, 
is obtained as 
\begin{eqnarray}
-\frac{1}{3}\str[AP(A)]=\bd s^2_{W_{2,4}} + \bd s^2_{T^{1,1}}\,. \label{coset-co}
\end{eqnarray}
Here $P$ denotes the projection that deletes the generators in (\ref{denominator}) 
from the Lie algebra $\mathfrak{g}$ of
\[
G = SL(2)_{\rm a} \times SL(2)_{\rm b} \times SO(2)_{\rm r} 
\times SU(2)_{\rm A} \times SU(2)_{\rm B}\times U(1)_{\rm R}\,.  
\] 

\medskip

It is also convenient to utilize another parametrization of $W_{2,4}$\,,  
in which two copies of Euclidean \AdS{2} are written in terms of the Poincar\'e coordinates.
The parametrization is
\begin{eqnarray}
g_{W_{2,4}}=\exp\left[-x_1 (L_2^1+L_0^1)+x_2 (L_2^2+L_0^2)+2 \chi K\right]
\exp\left[L_1^1\log z_1+L_1^2\log z_2\right]\,.
\label{PWpara}
\end{eqnarray}
In this parametrization, the generators $L^1_1$ and $L^{2}_1$ play a role of the dilatation generators 
of $\mathfrak{sl}(2)_{\rm a}$ and $\mathfrak{sl}(2)_{\rm b}$\,, respectively. 
Then the metric is given by
\begin{eqnarray}
\bd s^2_{W_{2,4}}=-\frac{1}{9}\left(\bd\chi+\frac{\bd x_1}{z_1}+\frac{\bd x_2}{z_2}\right)^2
+\frac{1}{6}\left(\frac{(\bd x_1)^2+(\bd z_1)^2}{(z_1)^2}+\frac{(\bd x_2)^2+(\bd z_2)^2}{(z_2)^2}\right)\,.
\end{eqnarray}

\medskip

The coset constructions introduced here will be the starting points in considering 
YB deformations of $W_{2,4}\times T^{1,1}$ in the following section.

\section{DTD models and YB deformations for PCM}

In this section, we will concentrate on a principal chiral model (PCM), 
instead of $W_{2,4}\times T^{1,1}$\,, 
in order to explain a relation between YB deformed PCMs and 
deformed T-dual (DTD) models \cite{HT-conjecture,BW-proof}. 

\subsection{DTD models for PCM}

The DTD models are realized by performing a non-abelian T-duality for the deformed PCM 
with a topological term, as explained below. 

\medskip 

Let us start from the classical action of PCM with a Lie group $G$\,, 
\begin{eqnarray}
S[g]=\frac{1}{2}\int_{-\infty}^{\infty}\!\!\bd\tau \int_0^{2\pi}\!\!\bd\sigma
\,\Tr \left[g^{-1}\partial_- gg^{-1}\partial_+ g\right]\,,
\end{eqnarray}
where $g$ is a group element of $G$\,. This system enjoys the global $G \times G$ symmetry like 
\[
g  \quad \longrightarrow  \quad g_{\rm L} \cdot g \cdot g_{\rm R}\,,
\]
where $g_{\rm L}$ and $g_{\rm R}$ are elements of the left and right global $G$'s\,, respectively.

\medskip

By gauging a subgroup $\widetilde{G}$ of a left global $G$\,, 
we obtain the following gauged action, 
\begin{eqnarray}
S[A,J,v] &=& \frac{1}{2}\int_{-\infty}^{\infty}\!\!\bd\tau \int_0^{2\pi}\!\!\bd\sigma
\,\Tr \left[(\tilde{A}_-+J_-)(\tilde{A}_++J_+)-v\tilde{F}_{+-}\right]\,,
\label{gaugePCM}
\end{eqnarray}
where the right-invariant current $J$ is defined as  
\[
J_\pm \equiv-\partial_\pm f \cdot f^{-1}\,, \qquad f\in G\,.
\]
Here $\tilde{F}_{+-}$ is the field strength of the gauge field $\tilde{A}_\pm$ that is defined as 
\begin{eqnarray}
\tilde{F}_{+-} \equiv \partial_+ \tilde{A}_--\partial_- \tilde{A}_+ -[\tilde{A}_+,\tilde{A}_-] 
\end{eqnarray}
and $v$ is a Lagrange multiplier taking a value of the ``dual algebra'' $\tilde{\mathfrak{g}}^*$\,. 
The generators for  $\tilde{\mathfrak{g}}$ and $\tilde{\mathfrak{g}}^*$ are described as 
\begin{eqnarray}
&&\tilde{T}_{\tilde{i}}~~: \qquad \mbox{the generators of}~\tilde{\mathfrak{g}} \nonumber \\ 
&& \tilde{T}^*_{\tilde{i}} ~: \qquad \mbox{the generators of}~\tilde{\mathfrak{g}}^{\ast} \nonumber 
\end{eqnarray}
and satisfy $\Tr[\tilde{T}_{\tilde{i}}\tilde{T}^*_{\tilde{j}}]=\delta_{\tilde{i}\tilde{j}}$\,. 
Note here that the range of the indices $\tilde{i},\tilde{j}\cdots$ is determined by the choice 
of the subgroup $\widetilde{G}$\,. 

\medskip 

The gauged action (\ref{gaugePCM}) is invariant under the following gauge transformation:\footnote{
If $\tilde{\mathfrak{g}}$ is not semi-simple, then $h\cdot v \cdot h^{-1}$ does not take 
the value in $\tilde{\mathfrak{g}}^{\ast}$ in general. 
For this point, see the footnote 11 of \cite{HT-conjecture}.}
\begin{eqnarray}
f\to h \cdot f\,,\quad \tilde{A}\to h \cdot \tilde{A} \cdot h^{-1} +{\rm d}h \cdot h^{-1}\,,
\quad v\to h \cdot v \cdot h^{-1}\,,\quad h\in \tilde{G}\,.
\end{eqnarray}
Integrating out the Lagrange multiplier $v$ gives rise to the zero curvature condition $\tilde{F}_{+-}=0$\,.
By taking a gauge $\tilde{A}_\pm=-\tilde{g}^{-1}\partial_\pm \tilde{g}$\,,
the original action with a group parametrization $g=\tilde{g}\cdot f$ can be reproduced. 
On the other hand, taking a variation with respect to $\tilde{A}$ corresponds 
to a non-abelian T-duality. 

\medskip

To obtain the action of DTD models, it is necessary to add the following topological term 
to the gauged Lagrangian (\ref{gaugePCM}) :
\begin{eqnarray}
\frac{\eta^{-1}}{2}\Tr\left[\tilde{A}_-\Omega \bigl(\tilde{A}_+ \bigr)\right]\,. 
\label{topoterm}
\end{eqnarray} 
Here $\eta$ is a constant real parameter which measures the deformation, and 
 $\Omega$ is a linear map from $\tilde{\mathfrak{g}}$ to 
the ``dual algebra'' $\tilde{\mathfrak{g}}^*$ satisfying the following condition: 
\begin{eqnarray}
\Omega \left({\rm ad}_x\, y \right) &=& {\rm ad}_x \Omega (y) - {\rm ad}_y \Omega (x)
\qquad  \left(~x\,,y\in\tilde{\mathfrak{g}}~\right)\,, \label{cocycle} \\ 
\Tr\left[x\, \Omega (y) \right]&=&-\Tr\left[\Omega(x)\,y\right]\,. 
\end{eqnarray}
Here the adjoint operation for the Lia algebra elements ad$_x\, y$ is defined as 
\[
{\rm ad}_x\,y \equiv [x,y]\,.
\]
The first condition (\ref{cocycle}) is called the cocycle condition. 

\medskip

Adding the topological term (\ref{topoterm}) corresponds to turning upon 
the following $B$-field \[
B\sim \Tr\left[\tilde{g}^{-1}d\tilde{g}\wedge \Omega( \tilde{g}^{-1}d \tilde{g})\right]
\] 
in the target space. 
The cocycle condition (\ref{cocycle}) requires that the induced $B$-field has to be closed.
In fact, $\tilde{A}=-\tilde{g}^{-1} \bd \tilde{g}$ satisfies the flatness condition   
$\bd\tilde{A}=\tilde{A}\wedge \tilde{A}$ and hence leads to the following expression:  
\begin{eqnarray}
\bd B&\sim &
\bd \Tr\left[\tilde{A}\wedge \Omega (\tilde{A})\right]
=-2\Tr\left[\tilde{A}\wedge \Omega(\tilde{A}\wedge \tilde{A})\right]\,.
\label{dB}
\end{eqnarray}
But here the right-hand side in (\ref{dB}) can be rewritten as follows:
\begin{eqnarray}
\Tr\left[\tilde{A}\wedge \Omega(\tilde{A}\wedge \tilde{A})\right]
&=&\frac{1}{2}\tilde{A}^i\wedge \tilde{A}^j\wedge\tilde{A}^k\,\Tr\left[\tilde{T}_i 
\Omega\,{\rm ad}_{\tilde{T}_j}\tilde{T}_k\right]\no \\
&=&\frac{1}{2}\tilde{A}^{\tilde{i}}\wedge \tilde{A}^{\tilde{j}}\wedge\tilde{A}^{\tilde{k}}\,\Tr\left[\tilde{T}_{\tilde{i}}({\rm ad}_{\tilde{T}_{\tilde{j}}} 
\Omega \tilde{T}_{\tilde{k}}-{\rm ad}_{\tilde{T}_{\tilde{k}}} \Omega \tilde{T}_{\tilde{j}})\right]\no \\
&=&
%2\Tr\left[\tilde{A}\wedge \tilde{A}\wedge \Omega(\tilde{A})\right]=
-2\Tr\left[\tilde{A}\wedge \Omega(\tilde{A}\wedge \tilde{A})\right]\,.
\label{dB2}
\end{eqnarray}
Note that the second equality in (\ref{dB2}) follows from the cocycle condition (\ref{cocycle}).
Thus the above new term (\ref{topoterm}) does not have effects 
on the classical dynamics of the gauged sigma models. 
%\begin{eqnarray}
%\bd B&\sim &
%\bd \Tr\left[\tilde{A}\wedge \Omega (\tilde{A})\right]
%=-2\Tr\left[\tilde{A}\wedge \Omega(\tilde{A}\wedge \tilde{A})\right]\,,
%\end{eqnarray}
%\begin{eqnarray}
%&=&-\Tr\left[\Omega(\tilde{A}\wedge \tilde{A})\wedge \tilde{A}
%+\tilde{A}\wedge \Omega(\tilde{A}\wedge \tilde{A})\right]\no \\
%&=&-2\Tr\left[\tilde{A}\wedge \Omega(\tilde{A}\wedge \tilde{A})\right]\no \\
%&=&-\tilde{A}^i\wedge \tilde{A}^j\wedge\tilde{A}^k\,\Tr\left[\tilde{T}_i 
%\Omega\,{\rm ad}_{\tilde{T}_j}\tilde{T}_k\right]\no \\
%&=&-\tilde{A}^{\tilde{i}}\wedge \tilde{A}^{\tilde{j}}\wedge\tilde{A}^{\tilde{k}}\,\Tr\left[\tilde{T}_{\tilde{i}}({\rm ad}_{\tilde{T}_{\tilde{j}}} 
%\Omega \tilde{T}_{\tilde{k}}-{\rm ad}_{\tilde{T}_{\tilde{k}}} \Omega \tilde{T}_{\tilde{j}})\right]\no \\
%&=&
%-4\Tr\left[\tilde{A}\wedge \tilde{A}\wedge \Omega(\tilde{A})\right]=
%4\Tr\left[\tilde{A}\wedge \Omega(\tilde{A}\wedge \tilde{A})\right]\,.
%\end{eqnarray}
Finally, we have shown that 
\[
\Tr\left[\tilde{A}\wedge \Omega(\tilde{A}\wedge \tilde{A})\right] = 0\,, \qquad \mbox{i.e.,} \quad 
\bd B =0\,.
\]
That is, the $B$-field is closed.

\medskip

Then the deformed gauged action is given by
\begin{eqnarray}
S&=&
%\frac{1}{2}\int_{-\infty}^{\infty}\bd\tau \int_0^{2\pi} \bd\sigma
%\,\Tr \left[(\tilde{A}_-+J_-)(\tilde{A}_++J_+)-v\tilde{F}_{+-}-\eta^{-1}\tilde{A}_- 
%\Omega (\tilde{A}_+)\right]\no \\&=&
\frac{1}{2}\int_{-\infty}^{\infty}\bd\tau \int_0^{2\pi} \bd\sigma
\,\Tr \left[\tilde{A}_-\tilde{\mathcal{O}}_+ (\tilde{A}_+) + \tilde{A}_-(\partial_+v +J_+)
-(\partial_- v-J_-)\tilde{A}_++J_-J_+\right]\,,
\label{dgaction}
\end{eqnarray}
where the operators $\tilde{\mathcal{O}}_\pm : \tilde{\mathfrak{g}}\to\tilde{\mathfrak{g}}^*$ 
are defined as
\begin{eqnarray}
\tilde{\mathcal{O}}_\pm \equiv \tilde{P}^T(1\pm{\rm ad}_v \mp\eta^{-1}\Omega)\tilde{P}\,.
\end{eqnarray}
Here the projection operators are defined as 
$\tilde{P} : \mathfrak{g}\to\tilde{\mathfrak{g}}$ and 
$\tilde{P}^T : \mathfrak{g}\to \tilde{\mathfrak{g}}^*$\,.
In particular, the operators $\tilde{\mathcal{O}}_\pm$ satisfy the following relations 
\begin{eqnarray}
\tilde{\mathcal{O}}_+^T=\tilde{\mathcal{O}}_-\,,\qquad
\tilde{\mathcal{O}}_\pm^{-1}\tilde{\mathcal{O}}_\pm=\tilde{P}\,,\qquad
 \tilde{\mathcal{O}}_\pm\tilde{\mathcal{O}}_\pm^{-1}=\tilde{P}^T\,,
\end{eqnarray}
where $\tilde{\mathcal{O}}_+^T$ is the transpose of the operator $\tilde{\mathcal{O}}_+$ 
that is defined through the relation 
\[
\Tr[x\,\tilde{\mathcal{O}}_+(y)]=\Tr[\tilde{\mathcal{O}}_+^T(x)\,y]\,.
\]
%The gauge symmetry is modified due to the existence of the deformation term.
%Indeed, it is necessary to add a shift transformation to the adjoint transformation of $v$ as 
%\begin{eqnarray}
%v\to h\cdot v \cdot h^{-1}+\eta^{-1}(........)\,.
%\end{eqnarray}

\medskip 

Then, taking a variation with respect to $\tilde{A}_\pm$ leads to the following expressions:
\begin{eqnarray}
\tilde{A}_+=-\tilde{\mathcal{O}}_+^{-1}(\partial_+ v+J_+)\,,\qquad
\tilde{A}_-=\tilde{\mathcal{O}}_-^{-1}(\partial_- v-J_-)\,.
\label{tA}
\end{eqnarray}
%From the definition of $\tilde{A}_\pm$\,, we obtain that 
%\begin{eqnarray}
%\tilde{g}^{-1}\partial_+\tilde{g}\simeq\tilde{\mathcal{O}}_+^{-1}(\partial_+ v+J_+)\,,\qquad
%\tilde{g}^{-1}\partial_-\tilde{g}\simeq-\tilde{\mathcal{O}}_-^{-1}(\partial_- v-J_-)\,.
%\label{replace}
%\end{eqnarray}
By putting these $\tilde{A}_\pm$ into the deformed gauged action (\ref{dgaction}), 
the classical action of the DTD model is derived as 
\begin{eqnarray}
S_{\rm DTD}&=&\frac{1}{2}\int_{-\infty}^{\infty}d\tau \int_0^{2\pi}d\sigma
\,\Tr \left[J_-J_++(\partial_- v-J_-)\tilde{\mathcal{O}}_+^{-1}(\partial_+v+J_+)\right]\,.
\label{DTDPCM}
\end{eqnarray}
At this stage, the left global symmetry $\tilde{G}_L$ is broken
but the right global symmetry $\tilde{G}_R$ is still preserved.
It is worth noting that the Lagrange multiplier $v$ plays 
a role of the dual coordinates in the dual models.

\subsection{YB deformations of PCM}

In this subsection, we will introduce YB deformations of PCM with the homogeneous 
CYBE \cite{MY} by following the terminology of \cite{BW-proof}. 

\medskip 

The action of the YB deformed sigma models \cite{Klimcik,DMV1,MY} is given by
\begin{eqnarray}
S_{\rm YB}=\frac{1}{2}\int_{-\infty}^{\infty}\bd\tau \int_0^{2\pi} \bd\sigma
\,\Tr \left[g^{-1}\partial_- g\frac{1}{1-\eta R_{g}}g^{-1}\partial_+ g\right]\,,
\end{eqnarray}
where $g$ is a group element of a Lie group $G$\,.
The deformed action has the right global symmetry $G_{\rm R}$ 
but the left global symmetry $G_{\rm L}$ is broken.
The dressed $R$-operator $R_g$ is defined by 
\begin{equation} 
R_g(x) \equiv g^{-1}R(gxg^{-1})g\,, \qquad  x\in \mathfrak{g}\,.
\end{equation}
Here the linear operator $R : \mathfrak{g}\to \mathfrak{g}$ is skew-symmetric and
satisfies the homogeneous CYBE 
\begin{eqnarray}
[R(x),R(y)]-R([R(x),y]+[x,R(y)])=0\,,\qquad x\,,y\in\mathfrak{g}\,.
\label{CYBE}
\end{eqnarray}
When the Lie algebra $\mathfrak{g}$ has a non-degenerate invariant symmetric bilinear form,
the $R$-operator is associated with a skew-symmetric classical $r$-matrix 
\begin{eqnarray}
r &=& \sum_i a_i \wedge b_i  
\equiv \sum_i ( a_i \otimes b_i - b_i \otimes a_i ) \qquad 
\in \mathfrak{g}\otimes\mathfrak{g}\,, 
\end{eqnarray}
which satisfies the homogeneous CYBE (in the tensorial notation) 
\begin{eqnarray}
[r_{12},r_{13}]+[r_{12},r_{23}]+[r_{13},r_{23}]=0\,.
\end{eqnarray}
Here the following tensor notations are utilized 
\[
r_{12}= \sum_i a_i\otimes  b_i\otimes 1\,, \quad 
r_{23}=\sum_i1\otimes  a_i\otimes  b_i\,, \quad 
r_{13}=\sum_i a_i\otimes 1\otimes b_i\,.
\]

\medskip 

In the following, we are concerned with a constant skew-symmetric solution of the CYBE. 
Then there are some nice properties. First of all, from the generators included 
in the classical $r$-matrix, a subalgebra of $\mathfrak{g}$ can be determined. 
Namely, if the classical $r$-matrix is represented by 
\begin{eqnarray}
r=\frac{1}{2}r^{\tilde{i}\tilde{j}}\,\tilde{T}_{\tilde{i}}\wedge \tilde{T}_{\tilde{j}}\,,
\qquad r^{\tilde{i}\tilde{j}}=-r^{\tilde{j}\tilde{i}}\,,
%\qquad {\rm det}\, r^{\tilde{i}\tilde{j}} \neq 0\,.
\end{eqnarray}
then the generators $\tilde{T}_{\tilde{i}}$ span a subalgebra $\tilde{\mathfrak{g}} 
\subset \mathfrak{g}$\,. 
Secondly, the determinant of $r^{\tilde{i}\tilde{j}}$ does not vanish 
on the subspace $\tilde{\mathfrak{g}}\otimes \tilde{\mathfrak{g}}$  \cite{Guide}\,, 
\begin{equation}
\det\,r^{\tilde{i}\tilde{j}} \neq 0\,. 
\end{equation}
%[A guide to Quantum groups(Proposition 2.2.6)]. 
It is worth noting that the action of $R$-operator is written as 
\begin{eqnarray}
R(x)=r^{\tilde{i}\tilde{j}}\tilde{T}_{\tilde{i}}\,\Tr[\tilde{T}_{\tilde{j}} x]\,,\qquad x\in\mathfrak{g}
\,.
\end{eqnarray}
Furthermore,  it is easy to see that $\tilde{\mathfrak{g}}$ is a quasi-Frobenius algebra  
\cite{Guide,Stolin}
%[A guide to Quantum groups(Corollary 2.2.4)]. 
which is equipped with a non-degenerate 2-cocycle $\omega$ satisfying the cocycle condition
\begin{eqnarray}
\omega([x,y],z)+\omega([z,x],y)+\omega([y,z],x)=0\,, \qquad x,y,z\in\tilde{\mathfrak{g}}\,.
\end{eqnarray}
In fact the 2-cocycle $\omega$ can be explicitly constructed by using the inverse of the $R$-operator : 
\begin{eqnarray}
\omega&\equiv&\frac{1}{2}\omega^{\tilde{i}\tilde{j}}\,\tilde{T}^*_{\tilde{i}}\wedge \tilde{T}^*_{\tilde{j}}
\equiv \frac{1}{2} (r^{-1})^{\tilde{i}\tilde{j}}\,\tilde{T}^*_{\tilde{i}}\wedge \tilde{T}^*_{\tilde{j}}\,,\\
\omega(x,y)&\equiv&  \omega^{\tilde{i}\tilde{j}}\,\Tr[\,x\,\tilde{T}^*_{\tilde{i}}]\Tr[\,y\,\tilde{T}^*_{\tilde{j}}]\,, \qquad x,y\in\tilde{\mathfrak{g}}\,.
\end{eqnarray}
Since $\omega$ is non-degenerate on $\tilde{\mathfrak{g}}\otimes \tilde{\mathfrak{g}}$,
it gives the skew-symmetric linear map $\Omega : \tilde{\mathfrak{g}}\to \tilde{\mathfrak{g}}^*$ which is defined by
\begin{eqnarray}
\Omega(x) 
\equiv \omega^{\tilde{i}\tilde{j}}\,\tilde{T}^*_{\tilde{i}}\,
\Tr[\,x\,\tilde{T}^*_{\tilde{j}}]\,,
\quad x\in\tilde{\mathfrak{g}}\,.
\label{Omega}
\end{eqnarray}
Note that $R\circ \Omega$ and $\Omega\circ R$ are the identity operators   
on $\tilde{\mathfrak{g}}$ and $\tilde{\mathfrak{g}}^*$\,, respectively.
The operator $\Omega$ also satisfies the cocycle condition (\ref{cocycle}).
To show the equivalence between a YB sigma model with a $r$-matrix and the corresponding DTD model,
it is necessary to choose the above operator $\Omega$  in the DTD model.

\subsection{The equivalence between DTD models and YB sigma models}

To see the equivalence between two sigma models,
we require the following conditions :
\begin{eqnarray}
\frac{1}{1\mp\eta R_{\tilde{g}}} = 1-\tilde{\mathcal{O}}_\pm^{-1}\,,\qquad
-\frac{1}{1\mp\eta R_{g}}g^{-1}\partial_\pm g = {\rm Ad}_f^{-1}(\tilde{A}_\pm+J_\pm)\,.
\label{opeq}
\end{eqnarray}
The conditions imply the field redefinition of the Lagrange multiplier
\begin{eqnarray}
{\rm d} v &=& -(\tilde{P}^T-\tilde{O}_+)\tilde{g}^{-1}{\rm d}\tilde{g}
=(\tilde{P}^T-\tilde{O}_-)\tilde{g}^{-1}{\rm d}\tilde{g} \nonumber \\ 
&=& \tilde{P}^T({\rm ad}_v -\eta^{-1}\Omega)\tilde{g}^{-1}{\rm d}\tilde{g}
\,.
\label{Dv}
\end{eqnarray}
%\begin{eqnarray}
%\partial_+ v=-(\tilde{P}^T-\tilde{O}_+)\tilde{g}^{-1}\partial_+\tilde{g}\,,\qquad
%\partial_- v=(\tilde{P}^T-\tilde{O}_-)\tilde{g}^{-1}\partial_-\tilde{g}\,.
%\label{Dv}
%\end{eqnarray}
For the integrated form of $v$\,, see Appendix A.
In particular, if matrix elements of ${\rm ad}_v$ are non-trivial,
the conditions (\ref{opeq}) determine constant of integrations 
from integrating of the constraints (\ref{Dv})\,.
Using the conditions (\ref{opeq}) or (\ref{Dv}) and the expressions (\ref{tA}) of $\tilde{A}$\,, 
we can show that the YB deformed action is rewritten as
\begin{eqnarray}
S_{\rm YB}&=&\frac{1}{2}\int_{-\infty}^{\infty}{\rm d}\tau \int_0^{2\pi}{\rm d}\sigma
\,\Tr \left[(-\tilde{g}^{-1}\partial_- \tilde{g}+J_-)(\tilde{A}_++J_+)\right]\no \\
%&=&\frac{1}{2}\int_{-\infty}^{\infty}{\rm d}\tau \int_0^{2\pi}{\rm d}\sigma
%\,\Tr \biggl[J_-J_++\tilde{g}^{-1}\partial_- \tilde{g}\tilde{\mathcal{O}}_+^{-1}(\partial_+ v+J_+)\no \\
%&&\qquad\qquad\qquad\qquad\qquad\qquad-\tilde{g}^{-1}\partial_- \tilde{g}J_+
%-J_-\tilde{\mathcal{O}}_+^{-1}(\partial_+ v+J_+)\biggr]\no \\
%&=&\frac{1}{2}\int_{-\infty}^{\infty}{\rm d}\tau \int_0^{2\pi}{\rm d}\sigma
%\,\Tr \biggl[J_-J_++(\tilde{\mathcal{O}}_-^{-1}\partial_- v+\tilde{g}^{-1}\partial_- \tilde{g})(\partial_+ v+J_+)\no \\
%&&\qquad\qquad\qquad\qquad\qquad\qquad-\tilde{g}^{-1}\partial_- \tilde{g}J_+
%-J_-\tilde{\mathcal{O}}_+^{-1}(\partial_+ v+J_+)\biggr]\no \\
&=&S_{\rm DTD}+\frac{1}{2}\int_{-\infty}^{\infty}{\rm d}\tau \int_0^{2\pi}{\rm d}\sigma
\,\Tr \left[\tilde{g}^{-1}\partial_- \tilde{g}\partial_+ v\right]\no \\
&=&S_{\rm DTD}+\frac{1}{2}\int_{-\infty}^{\infty}{\rm d}\tau \int_0^{2\pi}{\rm d}\sigma
\,\Tr \left[\tilde{g}^{-1}\partial_- \tilde{g}({\rm ad}_v -\eta^{-1}\Omega)\tilde{g}^{-1}\partial_+\tilde{g}\right]\,.
\label{eq}
\end{eqnarray}
%The third equality above follows from the relation
%\begin{eqnarray}
%\tilde{O}_-^{-1}\partial_- v=\tilde{O}_-^{-1}\tilde{g}^{-1}\partial_-\tilde{g} 
%- \tilde{g}^{-1}\partial_-\tilde{g}\,,
%\end{eqnarray}
%which is derived from (\ref{Dv}). 
The second term on the right-hand side in (\ref{eq}) is a total derivative term.
To see this, let us show the closure of the form 
\[
\Tr\left[\tilde{g}^{-1}{\rm d}\tilde{g}\wedge 
{\rm ad}_v( \tilde{g}^{-1}{\rm d} \tilde{g})\right]\,. 
\] 
This property follows from the following relation: 
\[
{\rm d}\Tr\left[\tilde{g}^{-1}{\rm d}\tilde{g}\wedge 
\Omega( \tilde{g}^{-1}{\rm d} \tilde{g})\right]=0\,.
\]
In fact,
\begin{eqnarray}
{\rm d}\Tr\left[\tilde{A}\wedge {\rm ad}_v\tilde{A}\right]
%&=&\Tr\left[-\tilde{A}\wedge\tilde{A}\wedge {\rm ad}_v\tilde{A}
%-\tilde{A}\wedge {\rm d}(v\tilde{A}-\tilde{A}v)\right]
%\no \\
&=&\Tr\biggl[-\tilde{A}\wedge\tilde{A}\wedge {\rm ad}_v\tilde{A}
+\tilde{A}\wedge v\tilde{A}\wedge \tilde{A}-\tilde{A}\wedge\tilde{A}\wedge \tilde{A}v\no \\
&&\qquad\qquad\qquad-\tilde{A}\wedge {\rm d}v\wedge \tilde{A}
-\tilde{A}\wedge \tilde{A}\wedge {\rm d}v\biggr]\no \\
%&=&\Tr\biggl[-\tilde{A}\wedge\tilde{A}\wedge {\rm ad}_v\tilde{A}
%+\tilde{A}\wedge\tilde{A}\wedge {\rm ad}_v\tilde{A}-2\tilde{A}
%\wedge \tilde{A}\wedge {\rm d}v\biggr]\no \\
&=&-2\Tr\biggl[\tilde{A}\wedge \tilde{A}
\wedge ({\rm ad}_v +\eta^{-1}\Omega)\tilde{A}\biggr]=0\,. 
\label{3.27}
\end{eqnarray}
In the third equality of (\ref{3.27}), we have used the relation
\begin{eqnarray}
\Tr\biggl[\tilde{A}\wedge \tilde{A}\wedge v\tilde{A}\biggr]
=\Tr\biggl[\tilde{A}\wedge \tilde{A}\wedge\tilde{A}v\biggr]\,.
\end{eqnarray}
%Finally, the equations in (\ref{replace}) can be rewritten as
%\begin{eqnarray}
%\tilde{g}^{-1}\partial_+\tilde{g}&\simeq&
%\frac{1}{1-\eta R_{\tilde{g}}}(\tilde{g}^{-1}\partial_+ \tilde{g}-J_+)
%+J_+\,,\no \\
%\tilde{g}^{-1}\partial_-\tilde{g}&\simeq&
%\frac{1}{1+\eta R_{\tilde{g}}}(\tilde{g}^{-1}\partial_- \tilde{g}-J_-)
%+J_-\,.
%\end{eqnarray}

\subsection{An example: deformed \AdS{3}}

In this subsection, we will give an explicit calculation of a YB deformation and
the corresponding DTD model for \AdS{3}.

\subsubsection*{Coset construction of AdS$_3$}

Let us introduce matrix realizations of $\mathfrak{sl}(2,\mathbb{R})$ that are described in
\begin{eqnarray}
t_0=\frac{i}{2}\sigma_2\,,\quad t_1=\frac{1}{2}\sigma_1\,,\quad t_2=\frac{1}{2}\sigma_3 \,,
\end{eqnarray}
where $\sigma_i$ are  the Pauli matrices.
We choose a group parametrization of $g\in SL(2,\mathbb{R})$ as
\begin{eqnarray}
g=\exp[2x^+t_+]\exp[2(\log z)\,t_2]\exp[2x^-t_-]\,.
\label{gsl2}
\end{eqnarray}
Here elements $t_\pm$ of $\mathfrak{sl}(2,\mathbb{R})$ are defined as
\begin{eqnarray}
t_+=\frac{t_0+t_1}{\sqrt{2}}\,,\qquad t_-=\frac{t_0-t_1}{\sqrt{2}}\,.
\end{eqnarray}
The left invariant current is expanded by a basis $\{t_2,t_\pm\}$ of $\mathfrak{sl}(2,\mathbb{R})$ as
\begin{eqnarray}
g^{-1}{\rm d}g&=&\frac{2{\rm d}x^+}{z^2}t_++2\left({\rm d}x^-
+2\left(\frac{x^-}{z}\right)^2{\rm d}x^+-\frac{2x^-}{z}{\rm d}z\right)t_-\no \\
&&+\frac{2(-2x^-{\rm d}x^++z{\rm d}z)}{z^2}t_2\,.
\end{eqnarray}
Then the associated metric of \AdS{3} spacetime is given by
\begin{eqnarray}
{\rm d}s^2= \frac{-2{\rm d}x^+{\rm d}x^-+{\rm d}z^2}{z^2}\,.
\end{eqnarray}
The metric is a familiar coordinate system of \AdS{3} in the Poincar\'e patch.

\subsubsection*{YB deformation of AdS$_3$}

First of all, we consider a YB deformation with the following non-abelian $r$-matrix \cite{MY}:
\begin{eqnarray}
r=\frac{1}{2}r^{\tilde{i}\tilde{j}}\,\tilde{T}_{\tilde{i}}\wedge \tilde{T}_{\tilde{j}}=2t_2\wedge t_+\,,
\end{eqnarray}
where generators $\tilde{T}_{\tilde{i}}$ of $\mathfrak{sl}(2,\mathbb{R})$ form 
a subalgebra $\tilde{\mathfrak{g}}$ of $\mathfrak{sl}(2,\mathbb{R})$ that is spanned by
\begin{eqnarray}
\tilde{\mathfrak{g}}&=&{\rm span}_{\mathbb{R}}\{\tilde{T}_1,\tilde{T}_2\}=
{\rm span}_{\mathbb{R}}\{\sqrt{2}\,t_2,\sqrt{2}\,t_+\}\,.
\label{qFsl2}
\end{eqnarray}
The non-vanishing commutation relation of $\tilde{\mathfrak{g}}$ is
\begin{eqnarray}
[\tilde{T}_1,\tilde{T}_2]=\sqrt{2}\,\tilde{T}_2\,.
\end{eqnarray}
In this basis of $\tilde{\mathfrak{g}}$\,, an element $r^{\tilde{i}\tilde{j}}$ of $r$-matrix becomes
\begin{eqnarray}
r^{\tilde{i}\tilde{j}}=
\begin{pmatrix}
0 \,& 1 \\
-1 & 0 
\end{pmatrix}
\,,
\end{eqnarray}
which has non-vanishing determinant.
The actions of linear operator $R$ is  
\begin{eqnarray}
R(t_2)=-t_+\,,\qquad R(t_-)=-t_2\,,\qquad R(t_+)=0\,.
\end{eqnarray}
The deformed background is given by 
\begin{eqnarray}
\bd s^2&=& \frac{-2\bd x^+\bd x^-+dz^2}{z^2}-\frac{\eta^2 (\bd x^-)^2}{z^4}\,,\no \\
B&=&\frac{\eta}{z^3} \bd x^-\wedge \bd z\,.
\label{Sch}
\end{eqnarray}
This background is nothing but a three-dimensional 
Schr\"odinger spacetime \cite{Domenico,Son,BM}.
Note here that the $B$-field is given by a total derivative term. 
This YB deformation was originally done in \cite{MY}. 

\subsubsection*{DTD model for AdS$_3$}

Let us next perform a similar computation for the corresponding DTD model.
Suppose that a duality group is given by a subgroup $\tilde{G}$ of $SL(2,\mathbb{R})$
that is generated by the Lie algebra $\tilde{\mathfrak{g}}$ (\ref{qFsl2})\,.  
The ``dual algebra'' $\tilde{\mathfrak{g}}^*$ can be introduced 
for the algebra $\tilde{\mathfrak{g}}$\,.
By the definition of the dual algebra, the basis of $\tilde{\mathfrak{g}}$ is given by
\begin{eqnarray}
\tilde{\mathfrak{g}}^*&=&{\rm span}_{\mathbb{R}}\{\tilde{T}^*_1,\tilde{T}^*_2\}=
{\rm span}_{\mathbb{R}}\{\sqrt{2}\,t_2,-\sqrt{2}\,t_-\}\,.
\label{dualb}
\end{eqnarray}
In this basis of $\tilde{\mathfrak{g}}$\,, the $2$-cocycle $\omega$ is given by
\begin{eqnarray}
\omega=\frac{1}{2}\omega^{\tilde{i}\tilde{j}}\,\tilde{T}_{\tilde{i}}^*\wedge 
\tilde{T}_{\tilde{j}}^*=2t_2\wedge t_-\,. 
\label{omegaSch}
\end{eqnarray}
Here, with the inverse of the classical $r$-matrix, $\omega^{\tilde{i}\tilde{j}}$ can be expressed as  
\begin{eqnarray}
\omega^{\tilde{i}\tilde{j}}=(r^{-1})^{\tilde{i}\tilde{j}}=
\begin{pmatrix}
0 \,& -1\\
1 \,& 0
\end{pmatrix}
\,.
\end{eqnarray}
Because the $2$-cocycle $\omega$ satisfies the cocycle condition,
the algebra $\tilde{\mathfrak{g}}$ is quasi-Frobenius.

\medskip

The next is to decompose the group element $g$ (\ref{gsl2}) as
\begin{eqnarray}
g&=&\tilde{g} \cdot f\,,\no \\
\tilde{g}&=&\exp[2x^+t_+]\exp[2(\log z)\,t_2]\,,\qquad
f=\exp[2x^-t_-]\,,
\end{eqnarray}
where $\tilde{g}$ is an element of the duality group $\tilde{G}$\,.
Then we can read off the dual background from the action (\ref{DTDPCM}) 
of a DTD model with the $2$-cocycle (\ref{omegaSch}).
The resulting background is given by
\begin{eqnarray}
\bd s^2&=&\frac{\eta}{2}\left[\frac{-4\bd x^-\bd v_1}{\sqrt{2}-2\eta v_2} 
+ \frac{-2\eta (\bd x^-)^2+\eta (\bd v_2)^2}{1-2\eta v_2(\sqrt{2}-\eta v_2)}\right]\,,\no \\
B&=&\frac{\eta}{2}\left[\frac{-\sqrt{2}\eta \bd x^-}{1-2\eta v_2(\sqrt{2}-\eta v_2)}
-\frac{\bd v_1}{1-\sqrt{2}\eta v_2}\right]\wedge \bd v_2\,. \label{3.47}
\end{eqnarray}
Thus the dual background is described in terms of the coordinates $x^-\,, v_1\,,$ and $v_2$\,, 
where $v_1$ and $v_2$ are the components of the Lagrange multiplier $v$ written as  
$v = v_1\,\tilde{T}^*_1 + v_2\,\tilde{T}^*_2$\,.

\medskip

The remaining task is to remove $v_1$ and $v_2$ from the expression (\ref{3.47})\,. 
For this purpose, it is helpful to perform a field redefinition of the Lagrange multiplier $v$\,. 
The first thing to consider is a constraint for the derivative of 
$v$ coming from (\ref{Dv})\,, 
which is given by
\begin{eqnarray}
\partial_\pm v&=&\partial_\pm v_1\,\tilde{T}^*_1+\partial_\pm v_2\,\tilde{T}^*_2 \no \\
&=&\frac{\sqrt{2}(1-\sqrt{2}\eta v_2)\partial_\pm x^+}{\eta z^2}\,\tilde{T}^*_1 
+ \frac{\sqrt{2}(1-\sqrt{2}\eta v_2)\partial_\pm z}{\eta z}\,\tilde{T}^*_2\,.
\label{Schdv}
\end{eqnarray}
Then we obtain a set of the first-order differential equations:
\begin{eqnarray}
\partial_\pm v_1 &=& \frac{\sqrt{2}(1-\sqrt{2}\eta v_2)\partial_\pm x^+}{\eta z^2}\,, \no \\  
\partial_\pm v_2 &=& \frac{\sqrt{2}(1-\sqrt{2}\eta v_2)\partial_\pm z}{\eta z}\,. 
\end{eqnarray}
By solving this system, $v$ can be determined as 
\begin{eqnarray}
v=(a_1-2a_2\,x^+)\tilde{T}^*_1+\left(\frac{1}{\sqrt{2}\eta}+a_2\,z^2\right) \tilde{T}^*_2\,,
\end{eqnarray}
where $a_1$ and $a_2$ are integration of constants.
The constraints (\ref{opeq}) fix the value of $a_2$ as 
\[
a_2=-\frac{1}{\sqrt{2}\eta}\,,
\] 
but $a_1$ cannot be determined. 
Thus we obtain the field redefinitions of the Lagrange multiplier $v$ like
\begin{eqnarray}
v=\frac{\sqrt{2}}{\eta}\left(
x^+ + \frac{\eta}{\sqrt{2}}\,a_1\right) \tilde{T}^*_1 
+ \frac{1-z^2}{\sqrt{2}\eta} \tilde{T}^*_2\,.
\label{intv}
\end{eqnarray}
After putting $v_{1,2}$ obtained from (\ref{intv}) into (\ref{3.47})\,,
the dual background can be rewritten as 
\begin{eqnarray}
\bd s^2&=& \frac{-2\bd x^+\bd x^-+\bd z^2}{z^2}-\frac{\eta^2 (\bd x^-)^2}{z^4}\,,\no \\
B&=&\frac{\eta}{z^3} \bd x^-\wedge \bd z+\frac{1}{\eta z}\bd x^+\wedge \bd z\,.
\end{eqnarray}
Thus the background is equivalent to a three-dimensional Schr\"odinger spacetime again,  
up to the total derivative term. Note that the ambiguity of the constant parameter $a_1$ 
has been absorbed into the shift symmetry for $x^+$\,.

\section{The equivalence for the  $W_{2,4}\times T^{1,1}$ case}

In this section, we show the equivalence between non-abelian T-dualities and 
YB deformations for the $W_{2,4}\times T^{1,1}$ case.  
This result indicates that the equivalence proposed for principal chiral models 
and symmetric coset sigma models should be valid even for non-symmetric coset cases. 

\subsection{DTD models for $W_{2,4}\times T^{1,1}$}

A unified picture of non-abelian T-dual models and homogeneous YB deformations 
has been constructed as a DTD model \cite{BW-proof}.
In the construction of \cite{BW-proof}, the symmetric coset structure 
was assumed so as to ensure the classical integrability. A remarkable point is that  
this picture is not restricted to the symmetric coset case but is still applicable to 
non-symmetric (non-integrable) cases as described below. 

\medskip

We start from the undeformed sigma model with target space $W_{2,4}\times T^{1,1}$\,.
The classical action is given by
\begin{eqnarray}
S=\frac{T}{3}\int_{-\infty}^{\infty}\bd \tau \int_{0}^{2\pi}\bd \sigma\,\ga^{\alpha\beta}\,
\str \left[g^{-1}\partial_\alpha g P(g^{-1}\partial_\beta g)\right]\,,
\label{uaction}
\end{eqnarray}
where $\ga^{\alpha\beta}={\rm diag}(-1,1)$ is the worldsheet metric and $T$ 
is the string tension. We will concentrate on the bosonic part hereafter 
and turn off the fermionic degrees of freedom. 
Hence $g$ is a representative of the coset (\ref{WTcoset})\,. 
It should be remarked that the projection $P$ has already been utilized 
in the coset construction (\ref{coset-co}) and this $P$ is not based 
on the grading structure unlike in the symmetric coset case. 

\medskip

As already explained in section 2, the $W_{2,4} \times T^{1,1}$ geometry 
is represented by a non-symmetric coset. However, due to the coset structure, 
it is still possible to construct the corresponding DTD models. 

\medskip 

Let us first construct a gauged action for a subgroup $\tilde{G}$\,.
The classical action is given by
\begin{eqnarray}
S=-\frac{T}{3}\int_{-\infty}^{\infty}\bd\tau \int_{0}^{2\pi}\bd\sigma\,
\str \left[(\tilde{A}_-+J_-)P_f(\tilde{A}_++J_+)
-v\,\tilde{F}_{+-}\right]\,,
\label{gaction1}
\end{eqnarray}
where $P_f={\rm Ad}_f\circ P\circ {\rm Ad}_f^{-1}$\,, $f\in G$
and $\tilde{F}_{+-}=\partial_+ \tilde{A}_--\partial_-\tilde{A}_+ -[\tilde{A}_+,\tilde{A}_-]$ is
the field strength of the gauge field $\tilde{A}$ for $\tilde{G}$\,.
$J_\pm=-\partial_\pm f f^{-1}$ is the right-invariant current and
$v$ is the Lagrange multiplier that takes values in the ``dual algebra'' $\tilde{\mathfrak{g}}$\,. 

\medskip 

To reproduce the original model, we integrate out the Lagrange multiplier $v$
that leads the flatness condition $\tilde{F}_{+-}=0$\,.
Then taking a pure gauge $\tilde{A}=-\tilde{g}^{-1}\bd \tilde{g}\,,\tilde{g}\in \tilde{G}$\,,
the gauged sigma model (\ref{gaction1}) is reduced to the original model with $g=\tilde{g}\cdot f$\,.

\medskip

The next task is to deform the gauged action (\ref{gaction1}) 
by adding the following topological term :
\begin{eqnarray}
\eta^{-1}\str\left[\tilde{A}_-\Omega\tilde{A}_+\right]\,.
\end{eqnarray}
Here the skew-symmetric operator $\Omega$ is the 2-cocycle satisfying the cocycle condition.
Then the action of the deformed gauged sigma model is given by
\begin{eqnarray}
S&=&-\frac{T}{3}\int_{-\infty}^{\infty}\bd\tau \int_{0}^{2\pi}\bd\sigma\,
\str \left[(\tilde{A}_-+J_-)P_f(\tilde{A}_++J_+)
-v\,\tilde{F}_{+-}-\eta^{-1}\tilde{A}_-\Omega\tilde{A}_+\right]\no \\
&=&-\frac{T}{3}\int_{-\infty}^{\infty}\bd\tau \int_{0}^{2\pi}\bd\sigma\,
\str \biggl[J_-P_fJ_++\tilde{A}_-\tilde{\mathcal{O}}_+\tilde{A}_+ \no \\ 
&&\qquad\qquad\qquad\qquad\qquad\qquad+\tilde{A}_-(\partial_+v +P_f(J_+))-(\partial_- v-P_f(J_-))\tilde{A}_+
\biggr]
\,, \label{gaction3}
\end{eqnarray}
where we have introduced the following operators: 
\[
\tilde{\mathcal{O}}_\pm=\tilde{P}^T(P_f\pm{\rm ad}_v\mp\eta^{-1}\Omega)\tilde{P}\,.
\]
Then taking a variation  with respect to the gauge field $\tilde{A}_\pm$ leads to 
\begin{eqnarray}
\tilde{A}_+=-\tilde{\mathcal{O}}_+^{-1}(\partial_+v +P_f(J_+))\,,\qquad
\tilde{A}_-=\tilde{\mathcal{O}}_-^{-1}(\partial_-v -P_f(J_-))\,.
\label{gA}
\end{eqnarray}
Putting the expressions of $\tilde{A}_\pm$ in (\ref{gA}) to the gauged action (\ref{gaction3}),
we get the action of DTD models 
\begin{eqnarray}
S_{\rm DTD}=-\frac{T}{3}\int_{-\infty}^{\infty}\bd\tau \int_{0}^{2\pi}\bd\sigma\,
\str\left[J_-P_f J_++(\partial_- v-P_f(J_-))\tilde{\mathcal{O}}_+^{-1}(\partial_+ v+P_f(J_+))\right]\,.
\label{DTD}
\end{eqnarray}
When the gauged subgroup $\tilde{G}$ is abelian,
the resulting background obtained from the DTD model is a TsT transformed background 
\cite{HT-conjecture}.

\medskip 

We will show that the action (\ref{DTD}) is equivalent to the action of YB sigma model,  
as shown in the next subsection. 

\subsection{The equivalence between DTD models and YB sigma models}

The YB deformed action is given by
\begin{eqnarray}
S_{\rm YB}=-\frac{T}{3}\int_{-\infty}^{\infty}\bd\tau \int_{0}^{2\pi}\bd\sigma\,
\str\left[A_- P\circ \frac{1}{1-\eta R_g\circ P}A_+\right]\,,
\label{YBW42}
\end{eqnarray}
where $R_g(x)=g^{-1}R(gxg^{-1})g$\,.
To show the equivalence between DTD models and YB sigma models,
we demand the following conditions :
\begin{eqnarray}
\frac{1}{1\mp \eta R_{\tilde{g}}\circ P_f}&=&1-\tilde{\mathcal{O}}^{-1}_\pm \circ P_f\,, 
\label{copeq} \\
-\frac{1}{1\mp\eta R_g\circ P}g^{-1}\partial_\pm g&=&{\rm Ad}_f^{-1}
\bigl(\tilde{A}_\pm+J_\pm\bigr)\,,
\label{dceq}
\end{eqnarray}
where $\tilde{A}_\pm$ is defined in (\ref{gA})\,.
These conditions indicate the constraints for derivatives of Lagrange multiplier $v$ like
\begin{eqnarray}
\bd v=-(\tilde{P}^T\circ P_f-\tilde{\mathcal{O}}_+)\tilde{g}^{-1}\bd \tilde{g}
=(\tilde{P}^T\circ P_f-\tilde{\mathcal{O}}_-)\tilde{g}^{-1}\bd \tilde{g}\,.
\label{vcWT}
\end{eqnarray}
Then the YB sigma model action (\ref{YBW42}) can be rewritten as
\begin{eqnarray}
S_{\rm YB}&=&-\frac{T}{3}\int_{-\infty}^{\infty}d\tau \int_0^{2\pi}d\sigma
\,\str \left[(-\tilde{g}^{-1}\partial_- \tilde{g}+J_-)P_f(\tilde{A}_++J_+)\right]\no \\
%&=&-\frac{T}{3}\int_{-\infty}^{\infty}d\tau \int_0^{2\pi}d\sigma
%\,\str \biggl[J_-P_fJ_++\tilde{g}^{-1}\partial_- \tilde{g}P_f\tilde{\mathcal{O}}_+^{-1}
%(\partial_+ v+P_f(J_+))\no \\
%&&\qquad\qquad\qquad\qquad\qquad\qquad-\tilde{g}^{-1}\partial_- \tilde{g}P_f(J_+)
%-J_-\tilde{\mathcal{O}}_+^{-1}(\partial_+ v+P_f(J_+))\biggr]\no \\
%&=&-\frac{T}{3}\int_{-\infty}^{\infty}d\tau \int_0^{2\pi}d\sigma
%\,\str \biggl[J_-P_fJ_++(\tilde{\mathcal{O}}_-^{-1}\partial_- v
%+\tilde{g}^{-1}\partial_- \tilde{g})(\partial_+ v+P_f(J_+))\no \\
%&&\qquad\qquad\qquad\qquad\qquad\qquad-\tilde{g}^{-1}\partial_- \tilde{g}P_f(J_+)
%-J_-\tilde{\mathcal{O}}_+^{-1}(\partial_+ v+P_f(J_+))\biggr]\no \\
&=&S_{\rm DTD}-\frac{T}{3}\int_{-\infty}^{\infty}d\tau \int_0^{2\pi}d\sigma
\,\str \left[\tilde{g}^{-1}\partial_- \tilde{g}\partial_+ v\right]\no \\
&=&S_{\rm DTD}-\frac{T}{3}\int_{-\infty}^{\infty}d\tau \int_0^{2\pi}d\sigma
\,\str \left[\tilde{g}^{-1}\partial_- \tilde{g}({\rm ad}_v -\eta^{-1}\Omega)
\tilde{g}^{-1}\partial_+\tilde{g}\right]\,.
\end{eqnarray}
The second term is a total derivative as in the PCM case. Thus the two models are equivalent 
at classical level even for the $W_{2,4} \times T^{1,1}$ case (up to a total derivative).

\section{Examples}

In this section, we present examples of YB deformations of 
the $W_{2,4}\times T^{1,1}$ background and the associated DTD models. 

\subsection{The case of abelian $r$-matrices}

Let us first consider a YB deformation associated with an abelian $r$-matrix
\begin{eqnarray}
r=\frac{1}{2}L^1_0\wedge L^2_0\,, 
\label{ar}
\end{eqnarray}
which satisfies the homogeneous CYBE and the unimodularity condition.
This classical $r$-matrix (\ref{ar}) is composed of the generators of the following algebra: 
\begin{eqnarray}
\tilde{\mathfrak{g}}&=&{\rm span}_{\mathbb{R}}\{\tilde{T}_1,\tilde{T}_2\}=
{\rm span}_{\mathbb{R}}\{\sqrt{2}\,L^1_0\,,\sqrt{2}\,L^2_0\}\,.
\label{qFa}
\end{eqnarray}
The dual algebra $\tilde{\mathfrak{g}}^*$ is spanned as 
\begin{eqnarray}
\tilde{\mathfrak{g}}^*&=&{\rm span}_{\mathbb{R}}\{\tilde{T}^*_1,\tilde{T}^*_2\}=
{\rm span}_{\mathbb{R}}\{-\sqrt{2}\,L^1_0\,,-\sqrt{2}\,L^2_0\}\,.
\label{duala}
\end{eqnarray}
Then the associated $2$-cocycle is given by  
\begin{eqnarray}
\omega=-8L^1_0\wedge L^2_0\,,
\label{aomega}
\end{eqnarray}
and satisfies the cocycle condition.

\medskip

The resulting YB deformed metric with $B$-field is given by
\begin{eqnarray}
\bd s^2&=&G(\hat{\eta})
\Bigl[-\frac{1}{9}(\bd \chi+\cosh y_1\,\bd\psi_1+\cosh y_2\,\bd\psi_2)^2
-\hat{\eta}^2\frac{\sinh^2 y_1\sinh^2 y_2}{324}\,\bd\chi^2\no \\
&&+\frac{1}{6}\sum_{i=1,2}\bigl(G(\hat{\eta})^{-1}\bd y_i^2+\sinh^2 y_i\,(\bd\psi_i)^2\bigr)
\Bigr]+\bd s^2_{T^{1,1}}\,, \nonumber \\
B_{\rm YB}&=&\hat{\eta}\,G(\hat{\eta})\Bigl[-\frac{\cosh y_2 \sinh^2y_1}{54}\,\bd\chi\wedge \bd\psi_1
+\frac{\cosh y_1\sinh^2y_2}{54}\,\bd\chi\wedge \bd \psi_2\no \\
&&-\left(\frac{\sinh^2 y_1\sinh^2 y_2}{36}-\frac{\sinh^2 y_1\cosh^2y_2
+\cosh^2y_1 \sinh^2y_2}{54}\right)\,\bd\psi_1\wedge \bd \psi_2
\Bigr]\,,
\label{abelianW}
\end{eqnarray}
where $\hat{\eta} \equiv \frac{3}{2}\eta$ and the scalar function $G(x)$ is defined as 
\begin{eqnarray}
G^{-1}(x) \equiv 1+x^2\left(\frac{\sinh^2 y_1\sinh^2 y_2}{36}
-\frac{\sinh^2 y_1\cosh^2y_2+\cosh^2y_1 \sinh^2y_2}{54}\right)\,.
\end{eqnarray}
Note here that the metric of the $T^{1,1}$ part has not been deformed. 

\medskip

The background is formally given by applying a double Wick rotation to 
the associated TsT transformed $T^{1,1}$ background  given in (3.16) and (3.17) 
of \cite{CMY}, up to the overall sign. 
Indeed, the deformed background (\ref{abelianW}) can be reproduced 
by applying the following TsT-transformation to the metric of $W_{2,4}$\,:
$1)$~perform a T-duality along the $\psi_1$-direction, 
$2)$~ shift $\psi_2$ like $\psi_2\to \psi_2-\frac{3}{2}\eta \psi_1$\,,
$3)$~perform a T-duality along the $\psi_1$-direction again. 

\medskip

To see the equivalence between this YB sigma model and the associated DTD model,
let us decompose a group element $g$ as follows:  
\begin{eqnarray}
g &=& \tilde{g}\cdot f\,,\no \\
\tilde{g} &=& \exp\left[\psi_1 L_0^1+\psi_2 L_0^2\right]\,,\quad
f = \exp\left[2\chi K\right]\exp\left[(y^1-i \pi)L_1^1+y^2 L_1^2\right]\cdot g_{T^{1,1}}\,.
\end{eqnarray}
Integrating out the constraint (\ref{vcWT}) under the decomposition leads to 
\begin{eqnarray}
v = \frac{4}{\eta}\, [(\psi_2+a_1) L_0^1 -(\psi_1+a_2)L_0^2\,]\,, \label{5.8}
\end{eqnarray}
where $a_{1\,,2}$ are constants of integration.
This $v$ also satisfies the requirements (\ref{copeq}) and (\ref{dceq}) for any values of $a_{1,2}$\,.
By using (\ref{5.8}), it is shown that the metrics of the two models are 
identical and the difference between the two $B$-fields is just a total derivative like 
\begin{eqnarray}
B_{\rm YB}-B_{\rm DTD}=-\frac{2}{3\eta}\,\bd\psi_1\wedge \bd\psi_2\,.
\end{eqnarray}
Thus the two models are equivalent up to this total derivative term.

\subsection{The case of non-abelian $r$-matrices}

The next example is a YB deformation with a non-abelian $r$-matrix, 
\begin{eqnarray}
r=2(L_0^1+L_2^1)\wedge L_1^1\,.
\label{ex2}
\end{eqnarray}
This is a solution of the homogeneous CYBE for $\mathfrak{sl}(2)$ and 
does not satisfy the unimodularity condition. 
The $r$-matrix (\ref{ex2}) is composed of the $\mathfrak{sl}(2)$ generators as  
\begin{eqnarray}
\tilde{\mathfrak{g}}&=&{\rm span}_{\mathbb{R}}\{\tilde{T}_1,\tilde{T}_2\}=
{\rm span}_{\mathbb{R}}\{L_0^1+L_2^1,\sqrt{2} L_1^1\}\,.
\label{qFna}
\end{eqnarray}
The dual algebra $\tilde{\mathfrak{g}}^*$ is spanned as 
\begin{eqnarray}
\tilde{\mathfrak{g}}^*&=&{\rm span}_{\mathbb{R}}\{\tilde{T}^*_1,\tilde{T}^*_2\}=
{\rm span}_{\mathbb{R}}\{L_2^1-L_0^1,\sqrt{2} L_1^1\}\,.
\label{dualna}
\end{eqnarray}
Then the associated $2$-cocycle is given by 
\begin{eqnarray}
\omega=-(L_2^1-L_0^1)\wedge L_1^1\,,
\label{naomega}
\end{eqnarray}
and satisfies the cocycle condition.

\medskip

In this case it is convenient to choose the parametrization (\ref{PWpara}).
The associated deformed background is given by
\begin{eqnarray}
\bd s^2&=&\frac{1}{6}\left(\frac{(\bd x_1)^2+(\bd z_1)^2}{(z_1)^2+\frac{\eta^2}{3}}
+\frac{(\bd x_2)^2+(\bd z_2)^2}{(z_2)^2}\right)\\
&&-\frac{1}{9}\left(1+\frac{2\eta^2}{3z_1^2+\eta^2}\right)
\left(\bd\chi+\frac{\bd x_1}{z_1}+\frac{\bd x_2}{z_2}\right)^2\\
&&+\frac{2\eta^2\bd x_1}{3z_1(3z_1^2+\eta^2)}
\left(\bd\chi+\frac{\bd x_1}{2z_1}+\frac{\bd x_2}{z_2}\right)
+\bd s^2_{T^{1,1}}\,, \\
B_{\rm YB}&=&\frac{\eta}{3(3z_1^2+\eta^2)}\bd z_1\wedge 
\left(\bd\chi -\frac{\bd x_1}{2z_1}+\frac{\bd x_2}{z_2}\right)\,.
\end{eqnarray}
Note here that the $r$-matrix (\ref{ex2}) does not satisfy the unimodular condition \cite{BW}. 
Hence the deformed background cannot be a solution of the usual type IIB supergravity, 
but should satisfy the generalized equations \cite{AFHRT} with appropriate completions.

\medskip

Finally, let us see the equivalence between the YB sigma model with the $r$-matrix (\ref{ex2}) 
and the associated DTD model. A group element $g$ can be parametrized as  
\begin{eqnarray}
g&=&\tilde{g}\cdot f\,,\no \\
\tilde{g}&=&\exp\left[-\frac{x_1}{z_1}(L^1_2+L_0^1)\right]\exp\left[L_1^1\log z_1\right]\,,\no \\
f&=&\exp\left[x_2 (L_2^2+L_0^2)+2 \chi K\right]\exp\left[L_1^2\log z_2\right]\cdot g_{T^{1,1}}\,.
\end{eqnarray}
Integrating out the constraint (\ref{vcWT}) under the decomposition leads to 
\begin{eqnarray}
v=\left(\frac{1}{2\eta}+a_1\,z_1\right)(L^1_2-L_0^1)+(\sqrt{2}a_1\,x_1+a_2)\sqrt{2}L_1^1\,,
\end{eqnarray}
where $a_{1\,,2}$ are constants of integration.
The requirements (\ref{copeq}), (\ref{dceq}) then determine $a_1=-\frac{1}{2\eta}$\,.
Using the expression of $v$\,, one can see that the metrics from the two models are identical 
and the difference between the NS-NS two-forms is just a total derivative: 
\begin{eqnarray}
B_{\rm YB}-B_{\rm DTD}=\frac{1}{6\eta z_1}\bd x_1\wedge \bd z_1\,.
\end{eqnarray}
Thus it has been shown that the two models are equivalent 
(up to the total derivative term).

\section{Conclusion and discussion}

In this note, we have studied Yang-Baxter deformations of type IIB string theory defined 
on the $W_{2,4}\times T^{1,1}$ spacetime with several examples of classical $r$-matrices 
satisfying the homogeneous CYBE. The result indicates that the Hoare-Tseytlin conjecture 
should be valid even for a non-symmetric coset (i.e.\ non-integrable) case. 
The analysis presented here was restricted to the metric (in the string frame) 
and NS-NS two-form. It would be very interesting to generalize to the supersymmetric case, 
though the Green-Schwarz string theory on the $W_{2,4}\times T^{1,1}$ background itself 
has not been constructed yet. 

\medskip 

As future directions, it would be interesting to study the conjecture relation 
for other examples of YB-deformations such as YB deformed Minkowski spacetime 
\cite{YB-Min,YB-Min2,PvT} and the deformed Nappi-Witten model \cite{NW,KY-NW}. 

\medskip 

As for the Yang-Baxter deformations with the mCYBE (including the $\eta$-deformation), 
some relations to non-abelian T-dualities have not been clarified yet. 
But it is well known that the $\eta$-deformation is equivalent 
to the $\lambda$-deformation \cite{lambda1,lambda2,lambda3} via the Poisson-Lie T-duality 
\cite{YB-lambda1,YB-lambda2} (For the recent progress on $\lambda$-deformations, 
see \cite{BTW,Lunin1,Lunin2} and \cite{BW}). 
It would also be nice to consider a generalization of this equivalence 
to non-symmetric coset cases like the $W_{2,4}\times T^{1,1}$ background. 

\medskip 

We hope that our work would provide a key to liberate YB deformations 
from the notion of integrability.

\subsection*{Acknowledgments}

We are very grateful to Hideki Kyono and Takuya Matsumoto for useful discussions. 
We would also like to thank Arkady Tseytlin, George Georgiou 
and Konstantinos Sfetsos for letting us know about the content of 
the works \cite{PZT} and \cite{Sfetsos-new}. 
We also grateful to Riccardo Borsato and Linus Wulff for the derivation of $v$ in Appendix A.
The work of J.S.\ was supported by the Japan Society for the Promotion of Science (JSPS).
The work of K.Y.\ was supported by the Supporting Program for Interaction-based Initiative 
Team Studies (SPIRITS) from Kyoto University and by a JSPS Grant-in-Aid for Scientific Research (C) 
No.\,15K05051. 
This work was also supported in part by the JSPS Japan-Russia Research Cooperative Program 
and the JSPS Japan-Hungary Research Cooperative Program.  

\appendix

\section*{Appendix}

\section{The integrated form of $v$}

We shall here derive the integrated form of the Lagrange multiplier $v$\,. 

\medskip

When a parametrization of $g$ is taken as
\begin{eqnarray}
g&=&\tilde{g}\cdot f\,,\qquad f\in G\,,\no\\
\tilde{g}& \equiv &\exp(RX)\in \widetilde{G}\,,\qquad X\in\tilde{\mathfrak{g}}^*\,,
\label{paratg}
\end{eqnarray}
the first condition in (\ref{opeq}) leads to the following form of $v\in\tilde{\mathfrak{g}}^*$~: 
\begin{eqnarray}
v=-\eta^{-1}\tilde{P}^{T}\frac{1-e^{-{{\rm ad}_{RX}}}}{{\rm ad}_{RX}}X\,.
\label{vex}
\end{eqnarray}
This form is given in \cite{BW-proof}, but for completeness we provide the derivation of this expression. 

\medskip

Let us rewrite the first condition in (\ref{opeq}) like
\begin{eqnarray}
\tilde{\mathcal{O}}_+^{-1}
&=&\frac{1}{\tilde{P}^T(1+{\rm ad}_v -\eta^{-1}\Omega)\tilde{P}}\no\\
&=&\tilde{P}\left(1-\frac{1}{1-\eta R_{\tilde{g}}}\right)\tilde{P}^T\no\\
&=&-\tilde{P}\frac{1}{1-\eta e^{-{\rm ad}_{RX}}\,R\,e^{{\rm ad}_{RX}}}(e^{-{\rm ad}_{RX}}(\eta^{-1}  \Omega) e^{{\rm ad}_{RX}})^{-1}\tilde{P}^T\no\\
&=&\frac{1}{\tilde{P}^T(1-\eta^{-1}  e^{-{\rm ad}_{RX}}\,\Omega\,e^{{\rm ad}_{RX}})\tilde{P}}\,.
\label{req}
\end{eqnarray}
Here the Campbell-Baker-Hausdorff formula leads to the following expression:  
\begin{eqnarray}
\tilde{P}^Te^{-{\rm ad}_{RX}}\,\Omega\,e^{{\rm ad}_{RX}}\tilde{P}
&=&\tilde{P}^T\left(\Omega-[{\rm ad}_{RX},\Omega]+\frac{1}{2}[{\rm ad}_{RX}\,,[{\rm ad}_{RX}\,,\Omega]]+\cdots\right)\tilde{P}\no\\
&=&\tilde{P}^T\sum_{n=0}^{\infty}\frac{1}{n!}\left((-{\rm ad}_{{\rm ad}_{RX}})^n\Omega\right)\tilde{P}\,.
\label{CBH}
\end{eqnarray}
Setting $x=RX\in \tilde{\mathfrak{g}}$ in the cocycle condition (\ref{cocycle}),
%\begin{eqnarray}
%\Omega \left({\rm ad}_x\, y \right) &=& {\rm ad}_x \Omega (y) - {\rm ad}_y \Omega (x)
%\qquad  \left(~x\,,y\in\tilde{\mathfrak{g}}~\right)\,, 
%\end{eqnarray}
we can obtain
\begin{eqnarray}
[{{\rm ad}_{RX}}\,, \Omega](y)={{\rm ad}_{y}}\,X=-{{\rm ad}_{X}}\,y\,,\qquad y\in\tilde{\mathfrak{g}}\,.
\label{cocycle2}
\end{eqnarray}
Using the expression (\ref{cocycle2}), (\ref{CBH}) can be rewritten as 
\begin{eqnarray}
\tilde{P}^Te^{-{\rm ad}_{RX}}\,\Omega\,e^{{\rm ad}_{RX}}\tilde{P}
%&=&\tilde{P}^T\left(\Omega+\sum_{n=1}^{\infty}\frac{1}{n!}(-1)^{n+1}\left(({\rm ad}_{{\rm ad}_{RX}})^{n-1}{\rm ad}_X\right)\right)\tilde{P}\no\\
&=&\tilde{P}^T\left(\Omega+\sum_{n=0}^{\infty}\frac{1}{(n+1)!}\left((-{\rm ad}_{{\rm ad}_{RX}})^{n}{\rm ad}_X\right)\right)\tilde{P}
\,.
\end{eqnarray}
Then the Jacobi identity for the Lie algebra $\mathfrak{g}$ enables us to derive the following relation: 
\begin{eqnarray}
\left({\rm ad}_{{\rm ad}_{RX}}{\rm ad}_X\right)(y)
&=&({\rm ad}_{RX}{\rm ad}_X-{\rm ad}_X{\rm ad}_{RX})(y)\no\\
&=&[RX,[X,y]]+[X,[y,RX]]\no\\
&=&[[RX,X],y]={\rm ad}_{{\rm ad}_{RX}X}(y)\,.
\end{eqnarray}
Hence, finally, we get the following expression :
\begin{eqnarray}
\tilde{P}^Te^{-{\rm ad}_{RX}}\,\Omega\,e^{{\rm ad}_{RX}}\tilde{P}
&=&\tilde{P}^T\left(\Omega+\sum_{n=0}^{\infty}\frac{1}{(n+1)!}(-1)^{n}{\rm ad}_{({\rm ad}_{RX})^n X}\right)\tilde{P}
\no \\
&=&\tilde{P}^{T}\left(\Omega -{\rm ad}_{-\tilde{P}^{T}\frac{1-e^{-{{\rm ad}_{RX}}}}{{\rm ad}_{RX}}X}\right)\tilde{P}\,. 
\label{CBH2}
\end{eqnarray}
Here we have ignored the deviation of $\frac{1-e^{-{{\rm ad}_{RX}}}}{{\rm ad}_{RX}}X$ from $\tilde{\mathfrak{g}}^*$ (For the detail, see the footnote 12 in \cite{HT-conjecture} ).

\medskip

Putting (\ref{CBH2}) into the last equation in (\ref{req}), the first condition in (\ref{opeq}) can eventually be written as 
\begin{eqnarray}
\frac{1}{\tilde{P}^T(1+{\rm ad}_v -\eta^{-1}\Omega)\tilde{P}}
=\frac{1}{\tilde{P}^T\left(1+{\rm ad}_{-\eta^{-1}\tilde{P}^{T}\frac{1-e^{-{{\rm ad}_{RX}}}}{{\rm ad}_{RX}}X}-\eta^{-1}\Omega\right)\tilde{P}}\,.
\end{eqnarray}
Thus we have obtained the integrated form of the Lagrange multiplier $v$ in  (\ref{vex})  
with the parametrization (\ref{paratg}).

\end{document}